# Mathematical modelling of an unstable bent flow using the selective frequency damping method

Alexander V. Proskurin

*Altai State Technical University, 656038, Russian Federation, Altai territory, Barnaul, Lenin prospect,*
k210@list.ru

**Abstract.** The selective frequency damping method was applied to a bent flow. The method was used in an adaptive formulation. The most dangerous frequency was determined by solving an eigenvalue problem. It was found that one of the patterns, steady-state or pulsating, may exist at some relatively high Reynolds numbers. The periodic flow occurs due to the instability of the steady-state flow. This numerical method is easy to use but requires a great deal of time for calculations.

## INTRODUCTION

In this paper, we consider the flow in a bent channel as shown in Figure 1. The channel consists of two parallel impermeable planes, curved so that the bending radius on the center line is equal to R. The distance between the planes is constant and equals to 2d. An incompressible viscous fluid flows under the action of constant pressure difference between the "in" and "out" regions. The Reynolds number $Re = Vd/\nu$ was introduced, where $V$ is the maximum velocity of the laminar flow, and $\nu$ is the fluid viscosity. The system of Navier-Stokes equations has the form

$$\frac{\partial \vec{V}}{\partial t} + (\vec{V}\nabla)\vec{V} = -\nabla p + \frac{1}{Re}\Delta \vec{V}, div\ \vec{V} = 0, \qquad (1)$$

where $\vec{V}$ is the velocity, $p$ is the pressure and $\vec{V} = 0$ for the condition to be valid on the channel walls. The Poiseuille parabolic profile is set at the channel entrance. The $\frac{\partial \vec{V}}{\partial x}$ condition is applied at the channel exit. Figure 2 shows meshes with 485(a) and 980(b) elements. The first mesh is used in calculations. The second mesh provides the convergence control. Velocity and pressure fields are calculated using the Nektar++ spectral-element framework[1]

The calculation is performed as a time-dependent problem: the initial state with zero velocity is set and the equations (1) is integrated. Depending on the Reynolds number, different flow patterns can be observed. For small Reynolds numbers, this is laminar flow or stationary vortices. If the Reynolds number is large enough, a pulsating flow may exist. Figure 3 shows the fields of the vertical velocity component at $Re = 700$ for these steady-state(a) and pulsating(b) modes. The appearance of the pulsating regime depends on the quality of calculations. If the time step decreases or a more detailed grid is used, the calculation converges to the steady flow at Re=700, and in the opposite case, the pulsating flow appears.

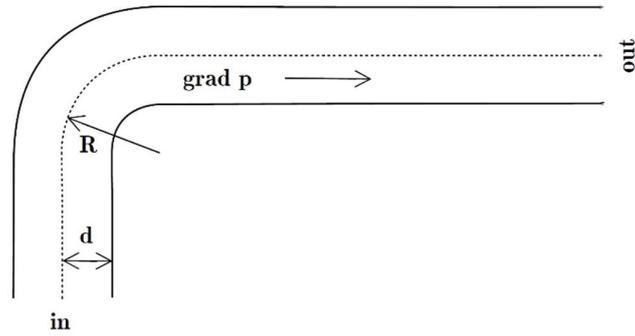

**FIGURE 1.** Flow sketch

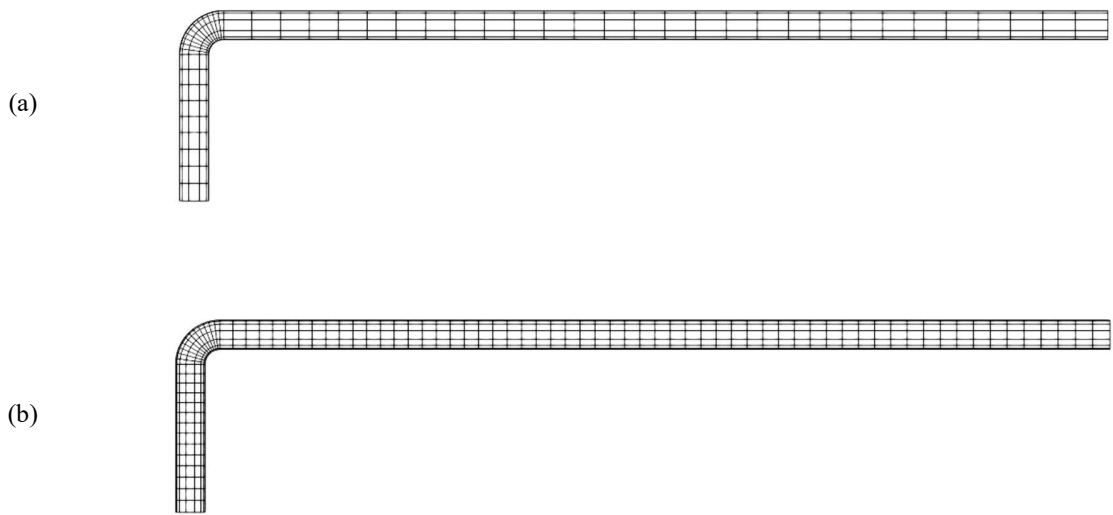

**FIGURE 2.** The mesh with 486 elements (a) and 980 elements (b)

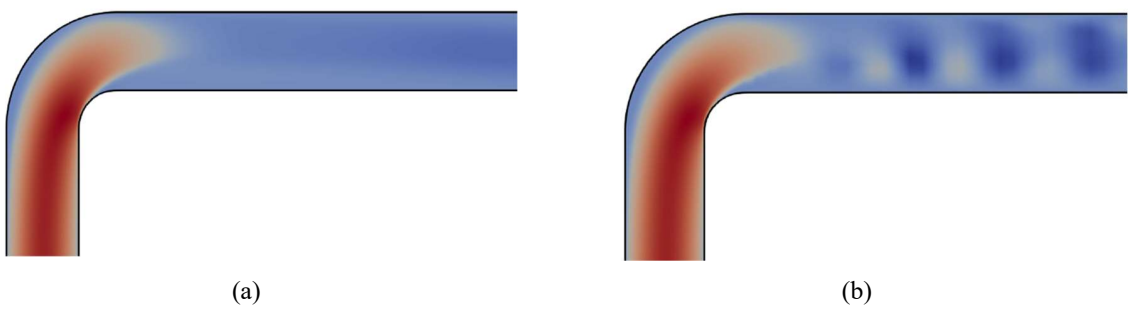

**FIGURE 3.** Vertical velocity fields at Re = 700 for steady-state(a) and pulsating (b) patterns

# NUMERICAL METHOD

In order to calculate the state flow at large Reynolds numbers, it is helpful to use the selective frequency damping method [2,3,4]. This method is implemented in the Nektar++ framework. By way of explanation (1) can be written in the space-discretized form

$$\dot{q} = F(q), \tag{2}$$

where $q$ is the flow variables and the dot over $q$ stands for the time derivation. Equation (2) can be modified

$$\dot{q} = F(q) - \chi(q - q_s) \tag{3}$$

where $q_s$ is the steady-state solution, and $\chi$ is the real positive value. Since the steady-state flow $q_s$ is unknown, it is replaced with the function

$$\bar{q}(t) = \int_{-\infty}^{t} exp\left(\frac{\tau-t}{\Delta}\right) q(\tau) d\tau \tag{4}$$

which is the filtered version of $q$. Expression (4) is a well-known exponential filter. This filter suppresses high frequency fluctuations in the flow. The $\Delta$ sets the threshold parameter. Using an equivalent differential form of (4) it is possible to write

$$\begin{cases} \dot{q} = F(q) - \chi(q - \bar{q}), \\ \dot{\bar{q}} = \frac{q-\bar{q}}{\Delta}. \end{cases} \tag{5}$$

A numerical splitting scheme is used to solve the problem (5). The scheme consists of two steps. In the first step, the nonlinear equation (2) is solved numerically. In discrete form, it is

$$\hat{q}_{n+1} = \Phi(q_n), \tag{6}$$

where $\Phi$ is the approximation of the function $F$. Without $F(q)$ the system (5) becomes the form

$$\begin{cases} \dot{q} = -\chi(q - \bar{q}) \\ \dot{\bar{q}} = \frac{q-\bar{q}}{\Delta} \end{cases} \text{ or } \begin{pmatrix} \dot{q} \\ \dot{\bar{q}} \end{pmatrix} = L \begin{pmatrix} q \\ \bar{q} \end{pmatrix}, L = \begin{pmatrix} -\chi I & \chi I \\ \frac{I}{\Delta} & -\frac{I}{\Delta} \end{pmatrix}. \tag{7}$$

A solution of equation (7) is

$$\begin{pmatrix} q(t+\delta t) \\ \bar{q}(t+\delta t) \end{pmatrix} = e^{L\delta t} \begin{pmatrix} q(t) \\ \bar{q}(t) \end{pmatrix}, \tag{8}$$

where $\delta t$ is the time step. Expression (8) determines the second step of the splitting scheme

$$\begin{pmatrix} q_{n+1} \\ \bar{q}_{n+1} \end{pmatrix} = e^{L\delta t} \begin{pmatrix} \hat{q}_{n+1} \\ \bar{q}_n \end{pmatrix} \tag{9}$$

To apply this numerical method, it is necessary to set the appropriate values for $\chi$ and $\Delta$. To do this, a one-dimensional model is imposed

$$u_{n+1} = e^{\lambda \delta t} u_n, \tag{10}$$

where $\lambda$ is the leading eigenvalue of the partially-converged solution $\bar{q}$. In this case, equation (9) can be written in the form

$$\begin{pmatrix} u_{n+1} \\ \bar{u}_{n+1} \end{pmatrix} = \exp\begin{pmatrix} -\chi & \chi \\ \frac{1}{\Delta} & -\frac{1}{\Delta} \end{pmatrix}\begin{pmatrix} \lambda & 0 \\ 0 & 1 \end{pmatrix}\begin{pmatrix} u_n \\ \bar{u}_n \end{pmatrix} = M(\lambda, \chi, \Delta)\begin{pmatrix} u_n \\ \bar{u}_n \end{pmatrix}. \tag{11}$$

The convergence of the scheme (6),(9) depends on the eigenvalues modules of $M$. If they are less than one, the scheme is stable. The optimal $\chi$ and $\Delta$ are determined from the minimum condition of these modules. This condition provides the fastest convergence of the scheme. Figure 4 shows the selective frequency damping numerical method workflow. During the program operation, the most dangerous modes were automatically determined by solving a linear eigenvalue problem and corresponding changes were made to the filter parameters.

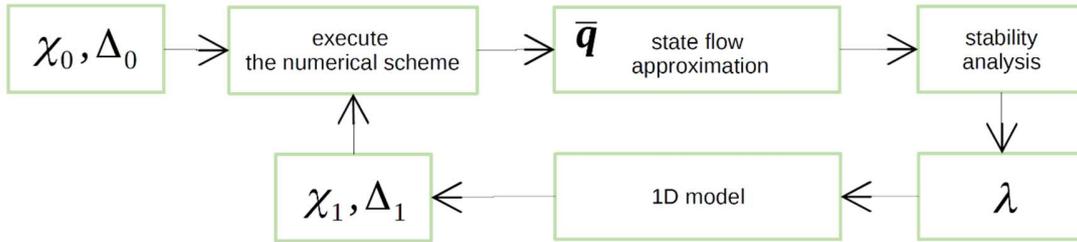

**FIGURE 4.** Calculations workflow

## RESULTS

The automatic selective frequency damping method and incompressible flow solver were implemented in the Nektar++ spectral-element framework [1]. Table 1 shows the values of the $y$-component of velocity in the outlet branch at the distance from the bend 20 and 40 (point 1 and 2, respectively) above the channel axis by 0.5, while the total length of the outlet branch is 60. The second column of the table shows the tolerance value that was set as a criterion for the method convergence. The third column shows the approximation order on each of the grid elements. The last column shows the calculation time on a computer with a 12-core AMD Ryzen Threadripper 1920X processor. For the values of the transverse velocity at points 1 and 2, there is convergence in two or three significant digits, and the difference in calculations on the different grids is less than one percent. For example, figure 5 shows the streamlines for pulsating (a) and steady-state (b) regimes where the Reynolds number equals 1200.

**Table 1.** *Convergence of vertical velocity component in the outlet branch at distance from the bend 20 and 40, distance from the outer plane is 0.5*

| Number of elements | Tolerance | Degree of approximation | Point 1 | Point 2 | Time of calculations (h:m:s) |
|---|---|---|---|---|---|
| 487 | 1E-06 | 5 | 0.00618925 | 0.00110740 | 2:57:35 |
| 487 | 1E-06 | 7 | 0.00429148 | 0.00113395 | 5:22:14 |
| 487 | 1E-06 | 10 | 0.00436140 | 0.00114256 | 5:40:46 |
| 487 | 1E-07 | 12 | 0.00439148 | 0.00114059 | 14:56:12 |
| 980 | 1E-04 | 12 | 0.004415 | 0.001151 | 15:24:56 |

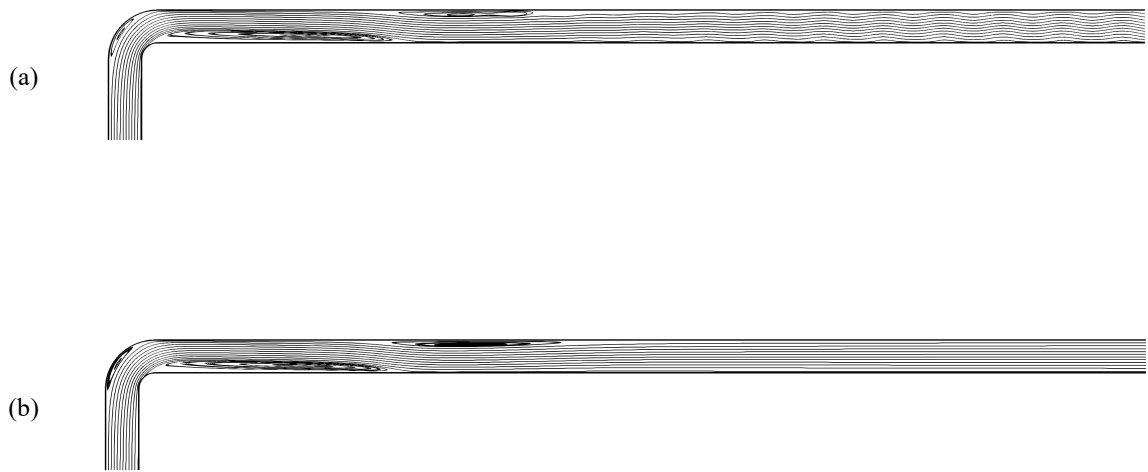

**FIGURE 5.** Streamlines at Re = 1200 for pulsating (a) and steady-state(b) patterns

# CONCLUSION

Using the adaptive selective frequency damping method, it was found that, for certain values of parameters, two solutions of the Navier-Stokes equations can exist in the channel: a steady-state and a periodical. The periodical flow occurs when the stability of the steady-state flow is lost under the influence of finite amplitude disturbances. These perturbations, for example, can be generated by the approximation errors of the numerical scheme.

This method is an alternative to the Newton method. This method is convenient because it does not need high quality initial approximations, as for the Newton method, and because it can be implemented in conjunction with existing non-stationary solvers, which in this case work on the "black box" principle. However, this method requires a considerable amount of time and does not converge for monotonically growing perturbations.